# Tunnelling Characteristics of Stone-Wales Defects in Monolayers of Sn and Group-V Elements

Pooja Jamdagni [1*], Ashok Kumar[2], Anil Thakur[3], Ravindra Pandey[4] and P.K.Ahluwalia[1]

[1]Department of Physics, Himachal Pradesh University, Shimla, H.P. India, 171005

[2]Centre for Physical Sciences, School of Basic and Applied Sciences, Central University of Punjab, Bathinda, India, 151001

[3] Department of Physics, Govt. P. G. College, Solan, H.P. India, 173212

4 Department of Physics, Michigan Technological University, Houghton, MI, 49931, USA, 49931

(June 22, 2017)

*Corresponding author:

Pooja Jamdagni (j.poojaa1228@gmail.com )

# Abstract

Topological defects in ultrathin layers are often formed during synthesis and processing, thereby, strongly influencing their electronic properties . In this paper, we investigate the role of Stone-Wales (SW) defects in modifying the electronic properties of the monolayers of Sn and group-V elements. The calculated results find the electronic properties of stanene (monolayer of Sn atoms) to be strongly dependent on the concentration of SW-defects e.g., defective stanene has nearly zero band gap (≈ 0.03 eV) for the defect concentration of 2.2 x $10^{13}$ $cm^{-2}$ which opens up to 0.2 eV for the defect concentration of 3.7 x $10^{13}$ $cm^{-2}$. In contrast, SW-defects appear to induce conduction states in the semiconducting monolayers of group-V elements. These conduction states act as channels for electron tunnelling, and the calculated tunnelling characteristics show the highest differential conductance for the negative bias with the asymmetric current-voltage characteristics. On the other hand, the highest differential conductance was found for the positive bias in stanene. Simulated STM topographical images of stanene and group-V monolayers show distinctly different features in terms of their cross-sectional views and distance-height profiles which can serve as fingerprints to identify the topological defects in the monolayers of group-IV and group-V elements in experiments.

# 1. Introduction

Due to outstanding properties and potential applications, atomically thin two-dimensional (2D) materials have received a great deal of attention in recent years [1-3]. Multilayer stanene that has been synthesized in a hexagonal lattice [4-5], crystallizes in a buckled structure with topological insulator properties [6-8]. Similarly, the group-V monolayers have gained particular attention [9-14]; e.g., arsenene, an atomic layer of As atoms, exhibits four different phases that switch into a topological insulator by an external electric field [15-16]; multilayer Sb shows a higher stability in β-phase [17] among the theoretically predicted four different phases [18-19]; 2D monolayer of Bi i.e. bismuthene, can be fabricated on a graphene substrate [20] with electronic structure that strongly depends on the spin-orbit coupling (SOC) factors [21-22].

2D materials are found to possess the so-called topological defects which appear due to regrouping of interatomic bonds without forming vacancies and /or addition of foreign impurity atoms, the simplest of such defect is Stone-Wales (SW) defect [23-27]. It has also been found that the energy barrier for the formation of SW-defects in buckled structures is lower than in graphene-like planner structures [28-29]. Moreover, SW-defects are found to be preferential adsorption centres for chemical functionalization [30]. SW-defects have also been observed in the atomic monolayers other than graphene [31-35].

In order to examine such topological defects in 2D materials, transmission electron microscope (TEM) [24] and scanning tunnelling microscope (STM) [36] are the commonly used experimental techniques. These techniques need to be supplemented with theory which provides an insight into origin and evolution of such defects [37-38]. One of the most useful formulations to describe tunnelling phenomenon in STM was proposed by Tersoff and Hamann [39-40] which uses Bardeen expression to calculate tunnelling current [41]. The wave function associated with the tip is assumed to be spherically symmetric (s-wave character) and tunnelling current can be calculated as convolution of local density of states (LDOS) of the tip and sample. LDOS in vacuum in terms of the LDOS of sample can be described by the Lang's approximation [42] where wave function in vacuum is allowed to decay exponentially.

The Bardeen, Tersoff and Hamann (BTH) model uses first-order perturbation theory to calculate tunnelling current that allows imaging of surface at atomic resolution [39-40]. The tunnelling characteristics using the BTH formalism have been successfully investigated previously for a wide variety of nanomaterials [38, 43-45]. In the present study, the BTH formalism together with density functional theory method have been used to investigate the current-voltage (I-V) characteristics and topography of SW-defects, in stanene and group-V monolayers. It has been found that SW-defects significantly alter the electronic structure and hence, the tunnelling characteristics of considered systems.

## 2. Computational Details

Electronic structure calculations were performed using density functional theory (DFT) as implemented in Vienna ab-initio simulation package (VASP) [46]. Generalized gradient approximation (GGA) within Perdew-Burke-Ernzerhof (PBE) parameterization is used to describe exchange-correlation functional. The van der Waals (vdW) interactions have been incorporated in the calculations by adding semi-empirical potential to the conventional Kohn-Sham DFT energy by using DFT-D2 method of Grimme [47]. We have also included the spin-orbit coupling (SOC) effects in our calculations as these were found to be important to describe electronic structure of considered systems [12, 15, 21-22]. A cut-off energy of 400 eV for the plane wave basis set and a Monkhorst–Pack mesh of (7 x 7 x 1) for Brillouin zone integration were employed. A 15 Å vacuum region in a (5 x 5 x 1) hexagonal supercell along z-direction ensures the modelling of a 2D material. All the structures are fully relaxed, with residual forces smaller than 0.01 eV/Å on each atom. Calculations were performed with varying SW-defect concentrations in the monolayers represented by the (5x5) and (6x6) periodic supercells,

3. **Results and Discussion**

The group-V elements exhibit variety of stable allotropic forms [15-19], though the graphene-like structure for group-V monolayers was considered. Table 1 lists the calculated structural properties of the monolayers considered. Our calculated values of buckling parameter ($\Delta$) are in excellent agreement with the previously reported values [Table 1]. Negative value of cohesive energy ($E_{coh}$) indicates the stability of given monolayers [48]. Note that the cohesive energy was obtained as: $\frac{E_T - nE_a}{n}$, where $E_T$ is the total energy of a supercell simulating a monolayer, $E_a$ is the energy of a free atom and n is the total number of atoms in a monolayer.

SW-defects in the monolayers of group-IV and group-V elements show distinctly different atomic reconstructions at the defective sites; e.g., in stanene, the dimer connecting the two pentagons through heptagon shows strong out-of-plan displacement of atoms as compared to antimonene (Figure 1). The relative formation energy ($E_{form}$) of defective monolayers decreases down the group from phosphorene (1.60 eV) to bismuthene (0.91 eV) [Table 1]. Note that $E_{form}$ was calculated as difference between the total energy of SW-defective and pristine monolayers. Formation energy of pristine monolayers is taken to be 0 eV as the reference energy. The monolayers which are energetically most stable have less tendency to form SW-defects and vice versa. $E_{form}$ of defective stanene is found to be dependent on the defect concentration, e.g., $E_{form}$ changes from 1.05 eV to 1.27 eV as we decrease the defect concentration. On the other hand, both relative formation energy and cohesive energy of defective group-V monolayers remain nearly the same on varying the defect concentration [Table 1].

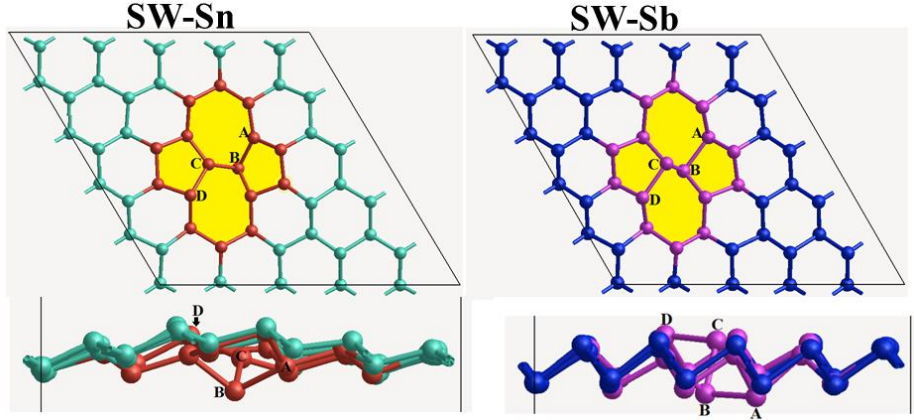

**Figure 1:** Top and side views of the relaxed structures of stanene (Sn) and antimonene (Sb) with single SW-defect.

**Table 1:** Buckling parameter (Δ), cohesive energy ($E_{coh}$), relative formation energy ($E_{form}$) and band gap ($E_{gap}$) of pristine and Stone-Wales (SW) defective monolayers in (5x5) and (6x6) periodic supercells

|  | **Pristine monolayers** | | | | **Defective monolayers** | | | | | | |
|---|---|---|---|---|---|---|---|---|---|---|---|
|  | Δ (Å) | $E_{coh}$ (eV) | $E_{gap}$ (eV) | | $E_{coh}$ (eV) | | $E_{form}$ (eV) | | $E_{gap}$ (eV) | | |
|  |  |  | without SOC | with SOC |  |  |  |  | without SOC | | with SOC |
|  | (5x5) | (5x5) | (5x5) | (5x5) | (5x5) | (6x6) | (5x5) | (6x6) | (5x5) | (6x6) | (5x5) |
| **Sn** | 0.85  0.86[a] | -3.24 | 0 | 0.07  0.074[a] | -3.22 | -3.23 | 1.05 | 1.27 | 0.20 | 0.03 | 0.20 |
| **P** | 1.23  1.24[b] | -5.18 | 1.97  1.97[b] | 1.97  1.97[b] | -5.15 | -5.16 | 1.60 | 1.57 | 1.53 | 1.58 | 1.53 |
| **As** | 1.39  1.40[b]  1.39[c] | -4.49 | 1.57  1.59[b]  1.62[c] | 1.41  1.81[b] | -4.46 | -4.47 | 1.32  1.27[f] | 1.28 | 1.25 | 1.37 | 1.23 |
| **Sb** | 1.64  1.65[b] | -3.88 | 1.17  1.26[b] | 0.98  1.00[b] | -3.85 | -3.86 | 1.06 | 1.01 | 1.00 | 1.05 | 0.82 |
| **Bi** | 1.73  1.71[b]  1.73[d]  1.74[e] | -3.60 | 0.51  0.55[b]  0.56[d]  0.55[e] | 0.50  0.43[b]  0.50[d]  0.51[e] | -3.58 | -3.58 | 0.91 | 0.94 | 0.57 | 0.53 | 0.32 |

[a]Ref. [6], [b]Ref. [12], [c]Ref. [15], [d]Ref. [21], [e]Ref. [22], [f]Ref [34]

## 3.1 Electronic Structure

Pristine stanene shows a strong spin-orbit coupling (SOC) effect which induces a band gap of 0.07 eV by opening the Dirac-like cone at K-point [Figure 2]. Single SW-defect in the (5x5) supercell of stanene with the defect concentrations of $2.2 \times 10^{13}$ cm$^{-2}$ introduces a band gap of 0.2 eV. An opening of the gap due to SW-defect is consistent with the previous theoretical studies performed on other group-IV monolayers such as silicene and germanene [2, 38]. Both valance band maximum (VBM) and conduction band minimum (CBM) are mainly composed of the out-of-plan p-orbitals.

On the other hand, an inclusion of the SOC effects in calculations find the group-V monolayers to be indirect band gap semiconductors [Table 1]. VBM mainly consists of the mixture of in-plan and out-of-plan p-orbitals while CBM is dominated by out-of-plan p-orbitals [Figure 2]. SW-defects introduce energy levels within the conduction band region that result into a reduction of band gap in defective monolayers. The defect energy levels mainly consist of $p_z$-orbitals of phosphorus atoms at the defective site. The valance bands near Fermi energy in the group-V monolayers show splitting due to the SOC effect. The splitting of bands increases in going from arsenene to bismuthene.

The electronic structure of the defective stanene is found to be sensitive to the defect concentration e.g., the band gap of the monolayer with SW-defect concentration of $2.2 \times 10^{13}$ cm$^{-2}$ is calculated to be 0.03 eV that changes to 0.2 eV for the defect concentration of $3.7 \times 10^{13}$ cm$^{-2}$. On the other hand, the band gap of the group-V monolayers remain almost unchanged with the defect concentration in the monolayers [Table 1]; e.g., on increasing the concentration of single SW-defect from 4.4 x $10^{13}$ cm$^{-2}$ to 6.8 x $10^{13}$ cm$^{-2}$ in phosphorene, the band gap value shows very small change from 1.58 eV to 1.53 eV. Note that the hexagonal supercell results in the band edges i.e. VBM and CBM, at K- for the (5x5) supercell case and at Γ-point for the (6x6) supercell case, that may result in different band dispersions for temporarily ordered defect configuration [28]. However, the value of the band gap remains the same irrespective of the position of band edges in the given monolayers.

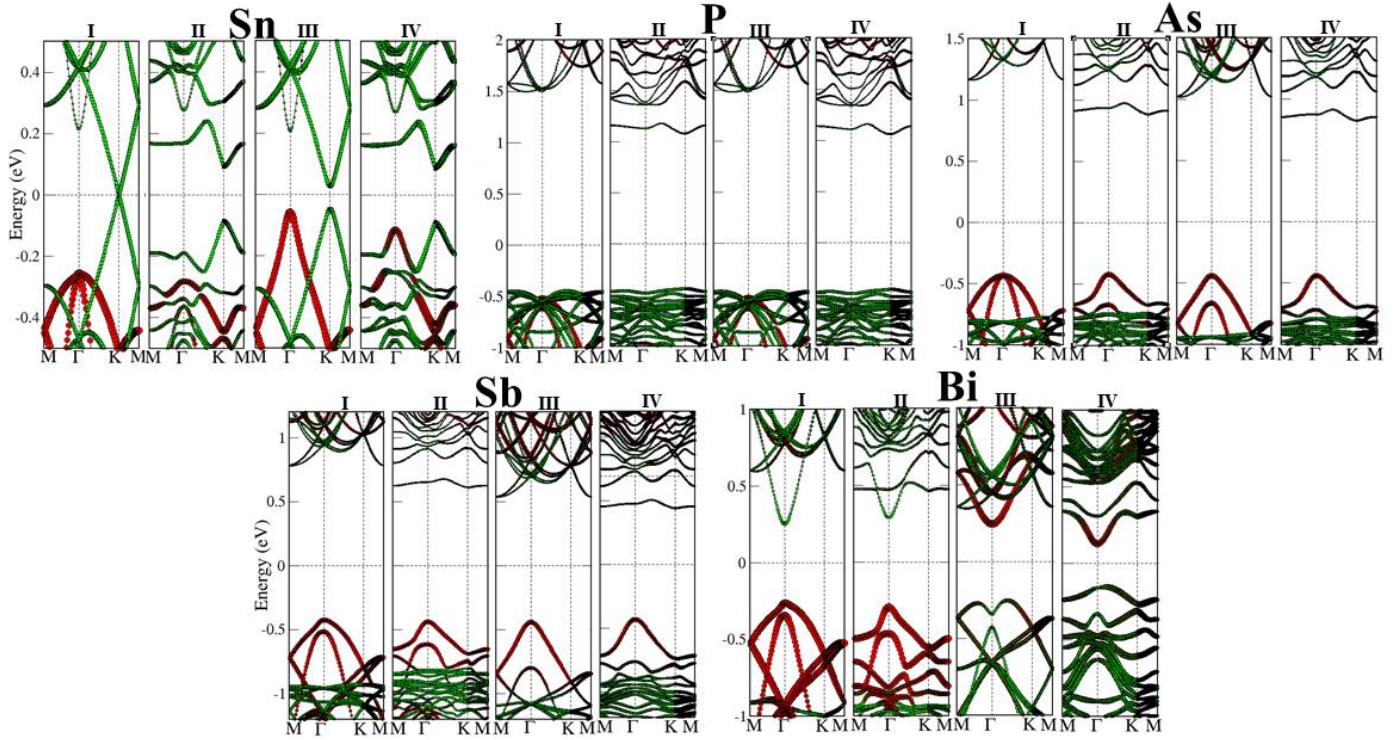

**Figure 2:** Projected electronic band structure of pristine and SW-defective monolayers of Sn and group-V elements. I and II represent the band structure of pristine and SW-defective monolayers without including spin-orbit coupling (SOC) effect while III and IV represent the corresponding band structure including SOC effect. Red and green colours represent the bands originating from $p_x + p_y$ and $p_z$ orbitals of respective atoms, respectively.

## 3.2 Tunnelling Characteristics

To study the tunnelling characteristics of the monolayers considered, we have used STM-like setup as shown in Figure 3(a). The $Au_{13}$ cluster is used to model the cap of the tip configuration in the STM setup. We use the BTH formalism to calculate electron tunnelling current [49] which is given by equation (1).

$$I = \frac{4\pi e}{\hbar} \int_{-\infty}^{+\infty} \rho_s(\varepsilon + \frac{eV}{2})\rho_t\left(\varepsilon - \frac{eV}{2}\right) e^{-2d\left(2\left(\frac{m}{\hbar^2}\right)(\varphi_{av}-\varepsilon)\right)^{\frac{1}{2}}} \left\{ \left[f\left(\varepsilon - \frac{eV}{2}\right)\right]\left[1 - \left[f\left(\varepsilon + \frac{eV}{2}\right)\right]\right] \right.$$

$$\left. - \left[f\left(\varepsilon + \frac{eV}{2}\right)\right]\left[1 - \left[f\left(\varepsilon - \frac{eV}{2}\right)\right]\right] \right\} \dots\dots\dots\dots\dots\dots (1)$$

Here, $\rho_s$ and $\rho_t$ are projected density of states (PDOS) of the sample (monolayer) and the tip ($Au_{13}$ cluster), respectively, obtained from DFT calculations. $\varepsilon$ is the injection energy of the tunnelling electron and $f$ is the Fermi distribution function. The distance '$d$' is separation between the tip and the sample. The value of the tip-sample separation of 5 Å is taken from the previously reported calculations [38, 45, 49-50]. The effective mass of electron (m) and average work function ($\varphi_{av}$,) of both sample and tip, are assumed to be constant in applied bias (± 1 V). The bias induced changes in the sample's DOS, which only appear at high enough applied bias [49-50], are not shown in the present study. Note that the BTH formalism is only applicable for small bias voltage $(V)$, $eV \ll \varphi_m$, where $\varphi_m$ is workfunction of the tip. Considering the average work function of metal, $\varphi_m \sim 4$ eV, the bias voltage limit ± 2 V is typically useful in the BTH formalism [51].

In order to understand the calculated tunnelling characteristics of the monolayers considered, we compared the DOS of sample (monolayer) and tip ($Au_{13}$ cluster) [Figure 3(b)].

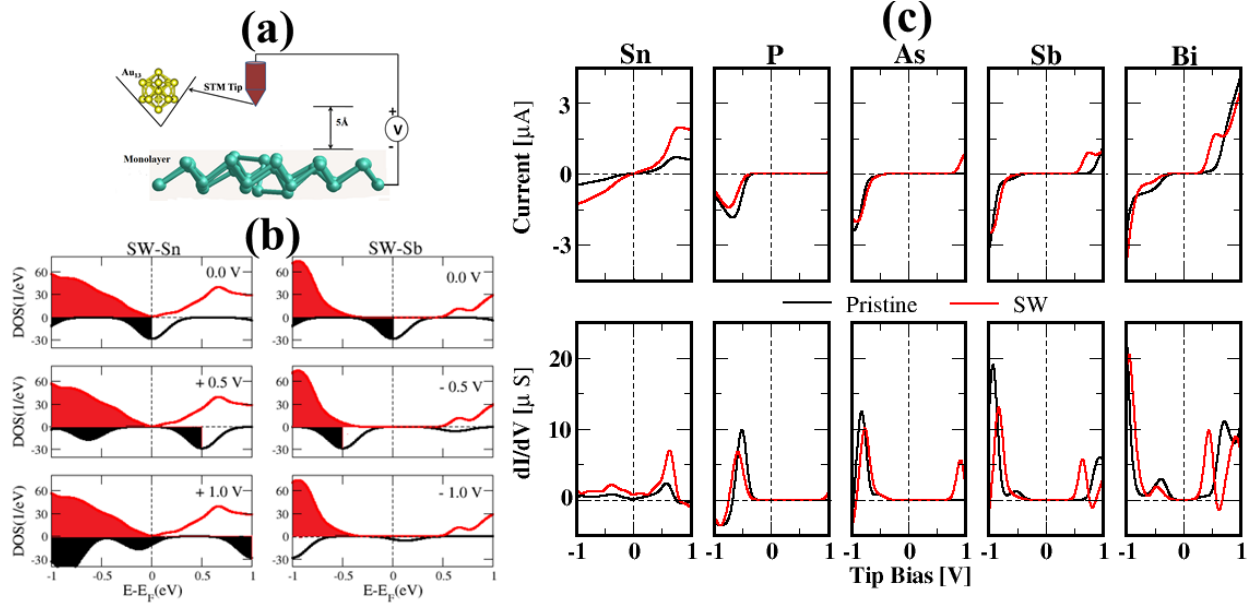

**Figure 3:** (a) STM-model setup, where distance 'd' between tip and sample is taken equal to 5 Å that enters in the exponential part in equation (1). (b) Density of states (DOS) of, stanene and antimonene with SW-defects (in-positive y-axis) and Au$_{13}$ tip (in negative y-axis), at different applied bias voltage where black and red shaded regions indicate occupied states of tip and monolayers, respectively. (c) Current-voltage characteristics and differential conductance ($dI/dV$) of pristine and SW-defective monolayers.

Note that tunnelling current is directly related to the convolution of the local density of states (LDOS) of tip and sample in the relevant regions. For example, none of the electrons of the tip or the sample see open channel to tunnel at zero bias in the defective stanene, so there can be no current. At +0.5 V bias voltage, the electrons within the energies of 0.5 $eV$ above the Fermi level see open channels to tunnel through sample that gives tunnelling current at +0.5 V [Figure 3(c)]. At the bias voltage of +1.0 V, the number of electrons available to tunnel through the sample remains nearly constant that leads to decrease in current [Figure 3(c)]. At very low bias (± 0.3 V), both pristine and defective-stanene show nearly linear current-voltage characteristics which is consistent with their semi-metallic nature.

On the other hand, the pristine group-V monolayers show a semiconducting behaviour. At higher positive bias in the defective group-V monolayers, the magnitude of the tunnelling current increases which is attributed to the available conduction channels facilitated by SW-induced electronic states near the Fermi level. Negative differential conductance (NDC) regions are also found at higher bias [Figure 3(c)] which may be attributed to the less number of available conduction channels to tunnel electrons in these systems. It is underlined that the choice of the tip-sample separation, which is taken to be 5 Å in our calculations, will affect only the magnitude of tunnelling current whereas the features of tunnelling spectra of considered systems remain the same [45, 49].

### 3.3 Simulated STM Images and Distance-Height Profiles

The STM topographical images can be obtained by applying a small bias voltage (V) between the sample and the tip that produces a tunnelling current whose density, j(r), is the simple extension [52] of the expression derived by Tersoff and Hamann [39-40].

$$j(r,V) \propto \rho_{STM}(r,V) \quad \text{------(2)}$$

$$\text{where} \quad \rho_{STM}(r,V) = \int_{E_F}^{E_F+eV} dE\, \rho(r,E) \quad \text{------(3)}$$

$$\text{and } \rho(r,E) = \sum_{n,k} |\psi_{nk}(r)|^2 \delta(E_{n,k} - E) \quad \text{------(4)}$$

where $\rho(r,E)$ is local density of states of the tip at $r$ and $\psi_{n,k}$ are the Kohn-Sham eigenstates obtained using density functional theory. Equation (3) describes the tunnelling from the occupied states of a sample to the tip. It is important to emphasize here that for the calculations of STM images, the tip states have been described by s-wave with constant density of states. The tunnelling matrix elements are considered to be independent of the lateral position of the tip for constant tip-to-sample distance and also independent of the bias voltage in narrow energy region [$E_F$, $E_{F+eV}$] [52].

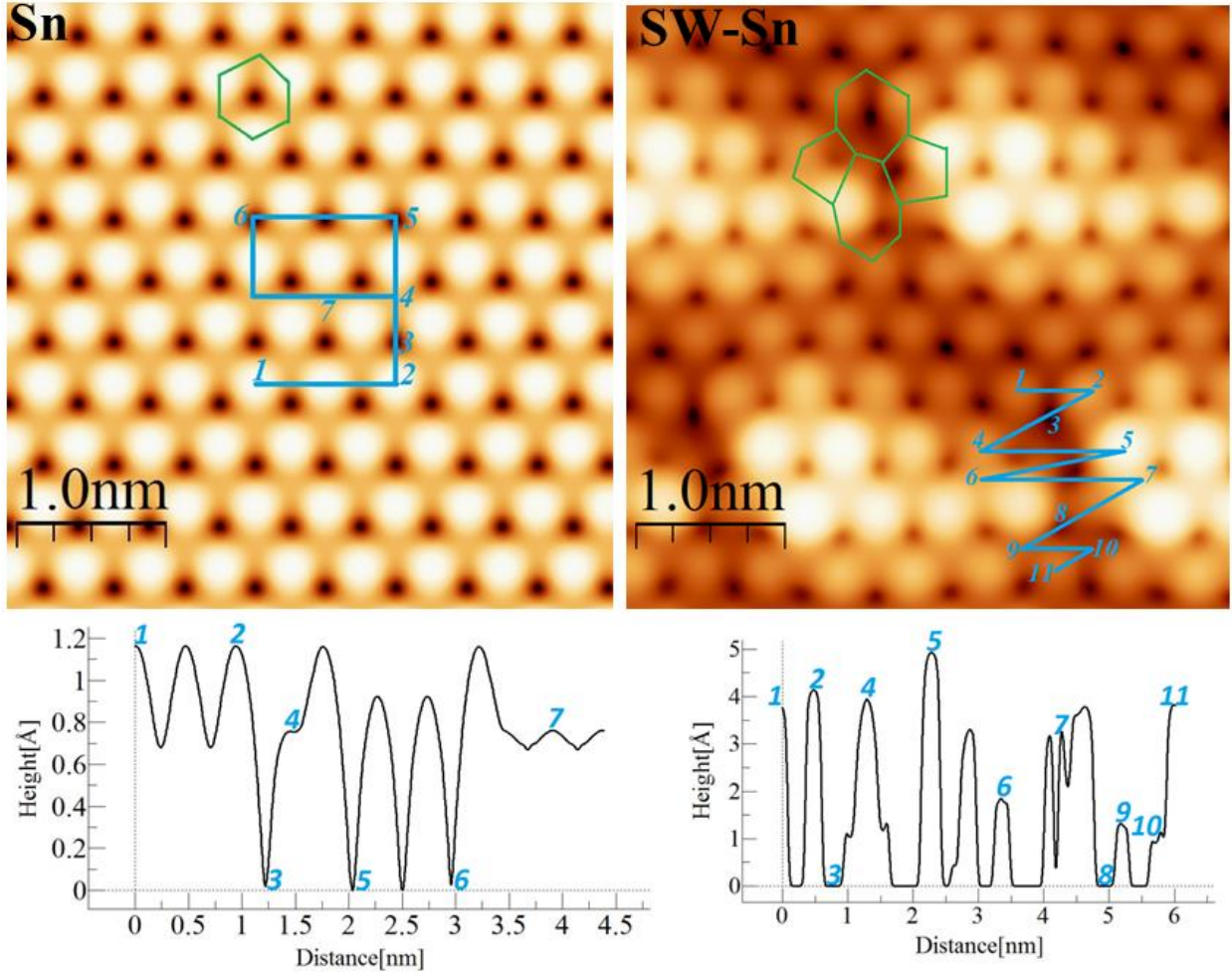

**Figure 4:** Simulated STM-images of pristine and SW-Sn with distance-height profile along the line shown in STM images, at bias voltage -1 V. Images are obtained at constant height corresponding to current imaging mode with isovalue varies as, (0 to 1.7) x $10^{-4}$ e/Å$^3$ in Sn and (0 to 4.8) x $10^{-4}$ e/Å$^3$ in SW-Sn. Brightest (darkest) regions in STM images are related with the highest (lowest) peak values in corresponding distance-height curve. The order of isovalue is kept same for all the images for the purpose of comparison.

Simulated STM images of pristine and defective monolayer of stanene at -1 V are displayed in Figures 4. STM image of pristine stanene shows white (brightest), light brown and dark brown (darkest) regions, respectively. The distance-height profile clearly indicates the three different height patterns corresponding to three different regions in STM image, e.g., positions marked as 2, 3 and 7 are recognized as upper atoms, characteristic holes and lower atoms, respectively, of stanene in honeycomb lattice.

The simulated STM image of the defective stanene also shows three different types of regions. The bright (white) region has been seen only for a few atoms which are at defective site and can be easily attributed to the strong out-of-plan reconstruction of atoms. The structure of SW-defects can be recognized using the distance-height scan [37]. The distance-height profile shows various atoms at different heights, e.g., positions marked as 1, 3 and 5 are recognized as one of the upper atom of heptagon, characteristic hole of heptagon and one of the upper atom of pentagon, respectively, of the defective stanene.

The STM images of all the group-V monolayers show similar features [Figure 5]. As a representative case, the detailed topography of pristine- and defective-antimonene monolayer is shown in Figure 5. Although it is difficult to distinguish between the characteristic hole region (marked as 5) and the lower buckled atom (marked as 3) in the STM image of pristine-antimonene at the bias of -1V, but the distance-height profile clearly shows a different height pattern for these regions [Figure 5].

SW-antimonene has distinctly different STM image than those seen for SW-stanene in terms of the cross-sectional view and the distance-height profile [Figures 4 and 5]. It is important to mention that, for the purpose of comparison, all the STM images are obtained using the same order of isosurface values (i.e. $10^{-4}$ e/Å$^3$) of local density of states, therefore, the colour contrast in various images may be different but the general features do not change. Our study concludes that, considering the topography of defected monolayer, STM may be useful to distinguish between SW-defects in the monolayers of group-IV and group-V elements.

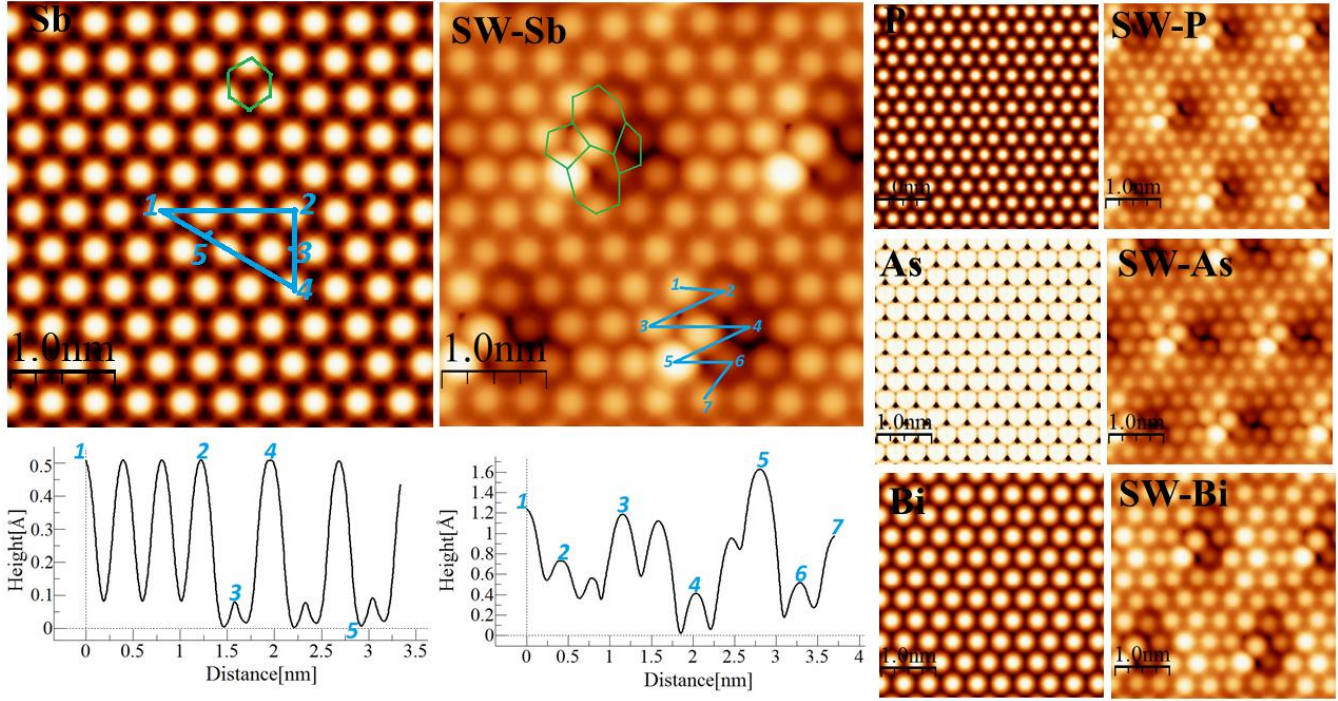

**Figure 5:** Simulated STM-images of pristine and SW-Sb with distance-height profile along the line shown in STM images at bias voltage -1 V. STM images of other group-V monolayer are also given. Images are obtained at constant height corresponding to current imaging mode with isovalue varies as (0 to 0.5) x $10^{-4}$ e/Å$^3$ and (0 to 1.6) x $10^{-4}$ e/Å$^3$ in Sb and SW-Sb, (0 to 0.3) x $10^{-4}$ e/Å$^3$ and (0 to 1.3) x $10^{-4}$ e/Å$^3$ in P and SW-P, (0 to 3.4) x $10^{-4}$ e/Å$^3$ and (0 to 1.6) x $10^{-4}$ e/Å$^3$ in As and SW-As, (0 to 0.6) x $10^{-4}$ e/Å$^3$ and (0 to 1.4) x $10^{-4}$ e/Å$^3$ in Bi and SW-Bi, respectively. Brightest (darkest) regions in STM images are related with the highest (lowest) peak values in corresponding distance-height curve. The order of isovalue is kept same for all the images for the purpose of comparison.

## 4. Summary

In summary, first principles calculations based on density functional theory are performed to investigate the Stone-Wales defects in monolayers of stanene and group-V elements. Stanene prefers the low-buckled honeycomb structures while the group-V monolayers prefer the high-buckled structures in their equilibrium configurations. The SOC effect induces ~ 70 meV band gap in stanene by opening up Dirac-cone whereas the band gaps of arsenene, antimonene and bismuthene are reduced by lifting the degeneracy at the valance band maximum. At very low bias, both pristine- and defective-stanene show ohmic current-voltage characteristics, while the group-V monolayers show zero tunnelling current. On increasing the bias voltage, for the defective group-V monolayers, the tunnelling current increases due to available defect-induced conduction channels. Since SW defect-sites can be identified using simulated STM images by scanning the distance-height profile, the present study may guide experimentalist in performing the STM measurements of defective monolayers.

## Acknowledgements

PJ is grateful to UGC New Delhi for providing financial assistance in the form of UGC-BSR junior research fellowship. Availability of CV Raman (DST, New Delhi, Govt. of India funded) cluster at Himachal Pradesh University, Shimla and RAMA High Performance Computing cluster at Michigan Technological University Houghton, USA are gratefully acknowledged for the present study.